\documentclass[11pt]{article}

\usepackage[english]{babel}
\usepackage[utf8]{inputenc}
\usepackage[usenames,dvipsnames]{color}

\usepackage[a4paper,left=2.5cm,right=2.5cm,top=2.5cm,bottom=2.5cm,bindingoffset=0mm]{geometry}

\usepackage{cite}

\usepackage{amsthm}
\usepackage{amssymb}
\usepackage{amsmath}

\usepackage{graphicx}

\usepackage[onehalfspacing]{setspace}

\newcommand{\tr}{\text{tr}}
\newcommand{\loss}{\mathcal{L}}

\begin{document}
\thispagestyle{empty}
\vspace*{-1.5cm}
\begin{flushright}
  {\small
  LMU-ASC 05/21\\
  MPP-2021-29
  }
\end{flushright}

\vspace{1.75cm}

\begin{center}
{\LARGE
Integrability ex machina}
\end{center}

\vspace{0.4cm}

\begin{center}
  Sven Krippendorf$^1$, Dieter L\"ust$^{1,2}$, Marc Syvaeri$^{1,2}$ 
\end{center}
\vspace{0.3cm}
\begin{center} 
\textit{$^{1}$\hspace{1pt} Arnold Sommerfeld Center for Theoretical Physics\\[1pt]
Ludwig-Maximilians-Universit\"at \\[1pt]
Theresienstra\ss e 37 \\[1pt]
80333 M\"unchen, Germany}
\\[1em]
\textit{$^{2}$\hspace{1pt} Max-Planck-Institut f\"ur Physik\\[1pt]
F\"ohringer Ring 6 \\[1pt]
80805   M\"unchen, Germany}
\end{center} 
\vspace{0.8cm}

%%%%%%%%%%%%%%%%%%%%%%%%%%%%%%%%%%%%%%%%%%%%%%%
%%%%%%%%%%%%%%%%%%%%%%%%%%%%%%%%%%%%%%%%%%%%%%%
%%%%%%%%%%%%%%%%%%%%%%%%%%%%%%%%%%%%%%%%%%%%%%%
%%%%%%%%%%%%%%%%%%%%%%%%%%%%%%%%%%%%%%%%%%%%%%%

\begin{abstract}
\noindent
Determining whether a dynamical system is integrable is generally a difficult task which is currently done on a case by case basis requiring large human input. Here we propose and test an automated method to search for the existence of relevant structures, the Lax pair and Lax connection respectively. By formulating this search as an optimization problem, we are able to identify appropriate structures via machine learning techniques. We test our method on standard systems of classical integrability and find that we can single out some integrable deformations of a system. Due to the ambiguity in defining a Lax pair our algorithm identifies novel Lax pairs which can be easily verified analytically.

\end{abstract}

\newpage
\tableofcontents

\section{Introduction}
Calculating quantities and determining the dynamics can be simplified in integrable systems due to additional integrals of motion. In analogy to symmetries which have associated conserved quantities integrability identifies (additional) symmetries in a system. 

Previously, some of us have shown how one can search for symmetries with the help of neural networks~\cite{Krippendorf:2020gny}. Neural networks have also been shown to identify the appropriate Hamiltonian~\cite{greydanus2019hamiltonian} and Lagrangian~\cite{cranmer2020lagrangian} from seeing phase space samples. From such networks one can determine the analytic formulae associated to the dynamics (e.g.~\cite{cranmer2020discovering}).
 
Here we are extending the search for such physically interesting functions to the case of integrability. Concretely, we are interested in identifying an appropriate Lax pair in the case of classical mechanics systems and to find a Lax connection in the context of field theory models~\cite{https://doi.org/10.1002/cpa.3160210503}. This search enables then answering whether a system is integrable. A simpler question seems to be to decide numerically whether a perturbation on a Hamiltonian is integrable.

The key to finding the appropriate Lax pair and connection is to formulate this search as a loss function which can be used to identify appropriate functions via machine learning. This is somewhat similar to other conditions such as the formulation of finding Ricci-flat metrics as an optimisation problem (cf.~for a recent machine learning aided search for Calabi-Yau and SU(3) structure metrics~\cite{Anderson:2020hux}). This search is inherently unsupervised, i.e.~it is not a trivial regression to known Lax pairs and connections. In a second step we are able to identify analytic formulae for Lax pairs which can be verified analytically.
To test our method we restrict ourselves here to standard examples of integrable systems which are widely discussed in the literature.

The rest of the paper is organized as follows: We firstly review conditions when a system can be called integrable (Section \ref{sec:review_integrability}). We then translate these conditions to optimization conditions suitable to be addressed via machine learning (Section \ref{sec:methods}). In Section \ref{sec:experiments} we present our experiments on finding Lax pairs and connections and Section~\ref{sec:interableornot} describes our experiments on identifying integrable perturbations. We conclude in Section \ref{sec:conclusions}.

%%%%%%%%%%%%%%%%%%%%%%%%%%%%%%%%%%%%%%%%%%%%%%%%%%%%%%%%%%%%%%%%%%%%%
\section{Lightning review of integrability}
\label{sec:review_integrability}
Our review on the key concepts and definitions of integrability is based on the review by Beisert~\cite{Beisertreview}.

Symmetries and associated  conserved quantities in physical systems simplify many calculations and are often a key factor in the analysis of systems. One concept here is the notion of integrability. Roughly speaking, a system is called integrable, if it has at least as many independent conserved quantities as degrees of freedom. For example the two-body problem in two dimensions is integrable due to the fact that it has four conserved quantities: total momentum, angular momentum (in the perpendicular direction) and the energy. An additional condition is the necessity that the conserved quantities are in involution, meaning that the Poisson brackets between them vanish, e.g. $\lbrace F_i, F_j\rbrace =0$ for all conserved quantities $F_i$ and $F_j$.

Studying classical integrability can be facilitated by introducing the concept of Lax pairs. A Lax pair is a pair of squared matrices $L$ and $M$ (mathematically speaking they are operators), which depend on the solution functions $p$ and $q$ which take values in phase-space. Additionally, the matrices depend on the spectral parameter $\lambda$, which becomes crucial for field theories. The defining property is, that they have to fulfill the equation
\begin{equation}
\begin{split} 
\label{eq:LaxPairCondition}
	\frac{\text{d}}{\text{dt}}L=\left[L,M\right]\,,
 \end{split} 
 \end{equation} 
and this equation has to be equivalent to the equations of motion. It is crucial to note, that these matrices are not uniquely defined, and therefore, many other choices are allowed as well. In the next step we can use the matrix $L$ to generate a tower of conserved quantities:
\begin{equation}
\begin{split} 
	F_k(\lambda) = \text{tr}\, L^k(\lambda)\,.
 \end{split}
 \end{equation}
To ensure that the conserved quantities are in involution $\lbrace F_k, F_l \rbrace = 0$, one can find the so called {\bf classical r-matrix} of the system. The defining equation is
\begin{equation}
\begin{split} 
	\lbrace L_1,L_2\rbrace = \left[r_{12},L_1\right] - \left[r_{21},L_2\right]\,,
 \end{split}\ \end{equation} 
where $L_1  := L \otimes 1$, $L_2  := 1 \otimes L$ and $r_{21}=P\left( r_{12}\right)$ with $P(\cdot)$ is the permutation operator between the two spaces.

An example is the harmonic oscillator with the equations of motion:
\begin{equation}\begin{split}
	\label{eq:eomosc}
	\dot{q} = p\,,\qquad \qquad \dot{p}	= -\omega^2 q\,.
 \end{split} \end{equation}
The Lax pair is not unique and we can easily identify two different families of consistent solutions:
\begin{equation}
 \begin{split}
	L_1= a\left(\begin{array}{cc}
	p &  b~\omega~q \\
	 \frac{\omega}{b} q & -p 
	\end{array} \right)\,,\qquad
	M_1= \left(\begin{array}{cc}
	0 &  \frac{b}{2} \omega  \\
- \frac{1}{2~b} \omega &  0
	\end{array} \right)\,,\\
	\\
	L_2= a\left(\begin{array}{cc}
	q &  \frac{1}{b~\omega} p \\
	 \frac{b}{w} p & -p 
	\end{array} \right)\,,\qquad
	M_2= \left(\begin{array}{cc}
	0 & - \frac{1}{2~b} \omega  \\
 \frac{2}{b} \omega &  0
	\end{array} \right)\,,\\
 \end{split}  \label{eq:HarmonicStandardSolution}
\end{equation}
where $a,b \in \mathbb{R}$. The solution known from the literature is $L_1$, $M_1$ with $a=b=1$. Then, the classical R-matrix becomes
\begin{equation} 
\begin{split}
	r_{12}=\frac{1}{q}\left(\begin{array}{cc}
	0 & 1\\
	0 & 0	
	\end{array}\right) \otimes \left(\begin{array}{cc}
	0 & 0\\
	1 & 0	
	\end{array}\right)	
	-
	\frac{1}{q}\left(\begin{array}{cc}
	0 & 0\\
	1 & 0	
	\end{array}\right) \otimes \left(\begin{array}{cc}
	0 & 1\\
	0 & 0	
	\end{array}\right)\,.
 \end{split} \label{eq:rmatrixcondition}\end{equation}
When we introduce the spectral parameter with $\tilde{L} = L + \lambda \mathbb{I}$, the conserved quantities become:
\begin{equation} \begin{split}
	F_1 ~=~& 2~ \lambda\,,\\
		F_2 ~=~& 2~ \lambda^2~ +~4~ H\,,\\
			F_3 ~=~& 2~ \lambda^3 ~+12~ \lambda~ H\,,\\
				F_4 ~=~& 2~ \lambda^4~ +24 ~ \lambda~ H~ +~ 4 ~H^2\,.
 \end{split} \end{equation}
This can be continued to all powers. We can see that the odd ones are trivial for $\lambda=0$ as they vanish, while the even ones depend on the Hamiltonian, and therefore are all not independent. Therefore, the only independent quantity is the Hamiltonian.\\

The next step is to extend this idea to the concept of {\bf field theories}, where infinitely many conserved quantities appear. For this, we are extending the concept of Lax pairs to the local concept of a {\bf Lax connection} $A\left(\lambda\right)$ which is a matrix-valued one form. In $1+1$ dimensions this reads as $A\left(\lambda\right) = A_x\left(\lambda\right) dx + A_t \left(\lambda\right) dt$.

This connection has to satisfy the flatness condition $dA = A \wedge A$, i.e.
\begin{equation} \begin{split}
	\dot{A}_x\left(\lambda\right) - A^{'}_t \left(\lambda\right) +\left[ A_x \left(\lambda\right), A_t \left(\lambda\right)\right] = 0\,,
 \end{split} \end{equation}
iff the equation of motion holds. This means that the flatness condition is only true when the equations of motions hold, so in particular it is not allowed to vanish.  Note that when we have a Lax connection  $A_t$, $A_x$, we can always reintroduce the spectral parameter by adding $\lambda \mathbb{I}$. In principle, the Lax pair $L$, $M$ can be obtained using the Lax connection as follows
\begin{equation} 
\begin{split}
	L\left(\lambda\right)=\vec{P}\exp\int^R_0 dx A_x \left(\lambda\right)\,,\qquad M\left(\lambda\right)=\left. A_t\left(\lambda\right)\right|_{x=0}\,,
 \end{split} 
\end{equation}
where we assumed that we have a compact space with length $R$, $\vec{P}$ is the path ordering operator.
For completeness, one would also have to show that they are in involution and therefore find the related classical r-matrix via the relationship
\begin{equation}
	\begin{split} 
		\lbrace L_1\left(\lambda_1\right),L_2\left(\lambda_2\right)\rbrace = 
		\left[r_{12}\left(\lambda_1,\lambda_2\right),
		L_1\left(\lambda_1\right)\otimes L_2\left(\lambda_2\right)\right]\,.
	\end{split}
\end{equation}

\section{Integrability structures from optimization}
\label{sec:methods}

We are interested in finding the Lax pair, respectively the Lax connection for field theoretical models for  different physical systems just by using the equations of motion and numerical solutions to these equations. To find either Lax structures we formulate the search as an optimization problem. We solve this optimization problem using a neural network ansatz for $L$, $M$ and the $r$-matrices. The procedure for both is as follows:
\begin{enumerate}
	\item Sample data points which fulfill  the equations of motion. Note, that we need no analytical solution, i.e.~for the harmonic oscillator we sample $p,q \sim \mathcal{N}\left(0,1\right)$, and compute $\dot{q}, \dot{p}$ subsequently by using the equations of motion (c.f. eq (\ref{eq:eomosc})).
        In our numerical searches we use $\sim 10^5~-~10^6$ data points which can be easily extended to sampling new samples for every epoch.
	\item Choose an ansatz for the Lax pair $L$, $M$. We usually start with the idea of having polynomials up to first order, and if the network does not converge we choose  higher order polynomials. In general, we can use an arbitrary neural network which corresponds to approximating non-polynomial functions. Due to the fact that we are looking at standard examples of integrability in this proof of concept study and we are interested in finding analytical expressions, we decided to focus on simple ans\"atze. Given an ansatz we then optimize our network subject to the integrability loss described below.
	\item Finally, we check the  results, whether they are actually equivalent to the differential equations, and compare the conserved quantities to the powers of $L$.
	\item Afterwards, we can use the presented framework to find solutions for the $r$-matrices. It is also possible to search for them directly when searching for the Lax pair.
\end{enumerate}
We now turn to the design of our loss functions. They have to enable  that the neural network is forced to suitable matrices which satisfy the following two conditions:
\begin{enumerate}
	\item The equation
\begin{equation}
\begin{split} 
\label{eq:laxcondition}
\frac{\text{d}}{\text{dt}}L \left(\lambda,p,q\right)-\left[L\left(\lambda,p,q\right),M\left(\lambda,p,q\right)\right] =0\qquad\text{or}\qquad
\dot{A}_x\left(\lambda\right) - A^{'}_t \left(\lambda\right) +\left[ A_x \left(\lambda\right), A_t \left(\lambda\right)\right] = 0
\end{split}
\end{equation}	
	 must hold for all sampled data points.
	\item This equation must be equivalent to the differential equations which define the physical system.
\end{enumerate}
Due to the fact that the data set is under our complete control, we can evaluate the time derivatives using the chain rule
\begin{equation}\begin{split}
	\frac{\text{d}}{\text{dt}}L=\frac{\partial L}{\partial p}\dot{p}+\frac{\partial L}{\partial q}\dot{q}\,,
 \end{split} 
 \end{equation}
where derivatives $\frac{\partial L}{\partial p}$ and $\frac{\partial L}{\partial q}$ can be evaluated using auto-differentiation.

The first condition (cf. eq (\ref{eq:laxcondition})) can be formulated as a loss function by considering the absolute values of the components. Our choice is to use the mean squared error, i.e.
\begin{equation} 
\begin{split}
 \loss_{\rm{Lax}}=\left|\left| \dot{L}-\left[L,M\right]\right|\right|^2\,,
 \end{split} 
 \end{equation}
where this norm is applied to each matrix component.

To find a loss function which for the second condition, we proceed as follows. We utilize that the equations of motion contain a time derivative and can be written as 
\begin{equation} 
\begin{split}	
\dot{x}_i = f_i\left(x_i,\partial x_i, ...\right)\,,
\end{split} 
 \end{equation}
 where $x_i$ is any quantity with a time derivative. Having only one time equation, and knowing that any time derivation of a variable originates from $\frac{dL}{dt}$, we can assume three things:
\begin{enumerate}

\item $L$ is at most of first order in the variables, e.g. $L = A_k x_k + B$.
\item Every component of $\dot{L}$ must be proportional to one of the $\dot{x}_k$ -- or  it has to vanish.
\item $\dot{x}_k$  must be proportional to at least one element of $\dot{L}$.
\end{enumerate}
The first point simplifies the structure of our network enormously, the latter two points give us the structure of the loss function.
\begin{equation} \begin{split}
\loss_{\rm{L}}=\sum_{i,j} \min_k\left( ||c_{ijk} \dot{L} -  \dot{x}_k||^2,||\dot{L}_{ij}||^2\right) + \sum_{k} \min_{ij}\left( ||c_{ijk} \dot{L}_{ij} -  \dot{x}_k||^2\right)\,,
 \end{split}
  \end{equation}
where $c_{ijk}=\frac{\sum_{batch}\dot{L}_{ij}}{\sum_{batch} \dot{x}_k}$ are the constants of proportionality\footnote{See \cite{Anderson:2020hux} for a similar loss which appears in the implementation of a Monge-Amp\`{e}re equation to find Ricci-flat Calabi-Yau metrics.}. The same procedure can be done for the equivalence of the remaining terms, so for $\left[L,M\right]$:
\begin{equation} \begin{split}
	\loss_{\rm{LM}}=\sum_{i,j} \min_k\left( ||\tilde{c}_{ijk} 
\left[ L,M\right]_{ij}-  f_k||^2,||\left[ L,M\right]_{ij}||^2\right) + \sum_{k} \min_{ij}\left( ||\tilde{c}_{ijk}\left[ L,M\right]_{ij}-  f_k||^2\right)\,,
 \end{split} \end{equation}
where $\tilde{c}_{ijk}=\frac{\sum_{batch}\left[ L,M\right]_{ij}}{\sum_{batch} f_k}$. Note that the loss term $\loss_{\rm{LM}}$ is not mandatory because the loss term $\loss_{\rm{Lax}}$ already contains the term, but it facilitates training.\\
As the last step we have to prevent mode collapse. To achieve this we demand that the sum of absolute values of the all components of $A_{k}$  has to be greater than some positive number:
\begin{equation} 
\begin{split}
	\loss_{\text{MC}} =\max \left(1- \sum \left|A_{ij} \right| , 0 \right)\,.
 \end{split}
  \end{equation}

The total loss is
\begin{equation}
	\loss_{\text{Lax-pair}} = \alpha_1 \loss_{\rm{Lax}}+\alpha_2 \loss_{\rm{L}} + \alpha_3 \loss_{\rm{LM}} + \alpha_4 \loss_{\text{MC}}\,,
\end{equation}
  where in our experiments we set $\alpha_1=\alpha_2=1$, $\alpha_4=10$ and $\alpha_3$ to 1 or 0.
\subsection*{Field theory}
This approach carries over to the field theory side. Basically, we replace all $L$ with $A_x$ and all $M$ with $A_t$. Therefore the loss components becomes:
\begin{equation} 
\begin{split}
 \loss_{\rm Lax}=&\sum_{ij}\left|\left| \dot{A}_x - A^{'}_t  +\left[ A_x , A_t \right]\right|\right|^2\\
 \loss_{A_x}=&\sum_{i,j} \min_k\left( ||c_{ij} \dot{A}_{x,ij} -  \dot{x}_k||^2,||\dot{A}_{x,ij}||^2\right) + \sum_{k} \min_{ijk}\left( ||c_{ijk} \dot{A}_{x,ij} -  \dot{x}_k||^2\right)\\
 \loss_{A_t}=&\sum_{i,j} \min_k
 \left(
  ||c_{ijk} \left[
  - A^{'}_t  +\left[ A_x,A_t \right] \right]_{ij}-  f_k||^2,||\left[- A^{'}_t  +\left[ A_x,A_t\right] 
  \right]_{ij}||^2,
  \right)\\
  &+ \sum_{k} \min_{ij}\left( ||c_{ijk}\left[ - A^{'}_t  +\left[ A_x ,A_t\right]\right]_{ij}-  f_k||^2
  \right)\, .
 \end{split} 
 \end{equation}

As last step we also prevent mode collapse. Again we demand that the sum of absolute values of the all components of $A_{k}$  has to be greater than some positive number:
\begin{equation} \begin{split}
	\loss_{\text{MC}} =\max \left(1- \sum \left|A_k \right| , 0 \right)\,.
 \end{split} \end{equation}
 
 The total loss for the Lax connection is
 \begin{equation}
	\loss_{\text{Lax-connection}} = \alpha_1 \loss_{\rm Lax}+\alpha_2 \loss_{A_x} + \alpha_3 \loss_{A_t} + \alpha_4 \loss_{\text{MC}}\,.
\end{equation}
 
For complex matrices, the complex space is interpreted as an additional dimension, and therefore, the summation and the minima are over all these dimensions.

\subsection*{Classical R-matrices}
R-matrices can be learned in the same way as the Lax pairs. Due to the fact that the left side of the equation is already fixed for a matrix $L$, we can train the network using only MSE on the equation:
\begin{equation}
\begin{split} 
	\loss_{\rm{R}} = \| \lbrace L_1,L_2\rbrace -\left[r_{12},L_1\right] + \left[r_{21},L_2\right] \|^2\,.
\end{split}
\end{equation}
\subsection*{Linear combinations within the Lax pair}
For some systems the Lax pair results necessarily in a linear combination of the equations of motion. One way of extending the framework to such systems is to compute the loss for each equation of motion individually, i.e when having two equations of motion
\begin{equation}
	\begin{split} 
		\dot{x}_1 &= f_1\left(x_1,x_2\right) \,,\\
		\dot{x}_2 &= f_2\left(x_1,x_2\right) \,.
	\end{split}
\end{equation}
Here we start with $\loss_{\rm L}$ for $x_1$, while setting $\dot{x}_2$ to zero, then continue with $x_2$, while setting $\dot{x}_1=0$. Using this method, we avoid the appearance of a linear combination, and therefore, the procedure still works. As this workaround does not seem to be as straight-forward for $\loss_{\text{LM}}$, we use a redundancy within $\loss_{\text{Lax-pair}}$ and therefore, we can simply set $\alpha_3=0$. 

We use this method only for the principle chiral model and for the distinction between integrable and non-integrable perturbations. In all the other cases, we use the regular loss $\loss_{\rm{Lax}}$.

\section{Experiments}
\label{sec:experiments}
Having set up our optimization, we now discuss several examples of the method.  First, we start with the harmonic oscillator. Then, we work our way to field theory using the Korteweg–de Vries equation (KdV) as a first example, and then go to two important physical examples: the Heisenberg model and the principal chiral model.

The most simple choice to build a neural network is to interpret every prefactor of the polynomial as an independent free parameter and train it using the framework of gradient-based optimization known from machine learning. We realized during the experiments that neural networks with such a small number of parameters (while having complicated constraints) is heavily dependent on the initialization of the network. To avoid these limitations, we decided to use a more sophisticated way on how to compute the prefactors.
For each element $a$ the neural network is trained to fit a vector $\vec{a}$ of $2n+1$ elements, where $n$ is an integer. Then, each prefactor is calculated using the following operator:
\begin{equation}
\begin{split} 
	a = \text{SoftMax}\left(\vec{a}\right)\cdot \vec{v}
	\qquad\text{with} \qquad 
	\vec{v}=	\left( \begin{matrix}
	n  \\
	\vdots\\
	-n
	\end{matrix}\right)\qquad\text{and} \qquad \text{SoftMax}\left(\vec{a}\right)_i=\frac{e^{a_i}}{\sum_k e^{a_k}}\,.
\end{split}
\end{equation}

This ensures that the neural network can faster change the prefactors and therefore results in a more stable training process. Note, that the prefactors are  limited to the range $\left[-n,n\right]$, and therefore, we usually choose $n=5$ which is large enough to not restrict the neural network in any way.

Additionally, we can use the values $\vec{a}$ to support the neural network to avoid finding linear combinations of $x_k$ within $L$. Assuming, we have multiple $x_1,\ldots,x_n$, we can define the prefactor $a_k$ like:
\begin{equation}
	a_k =  \sigma\left(\sum_i a_{k,i}-\sum_{j\neq k}\sum_{i}a_{j,i}\right)  ~\text{SoftMax}\left(\vec{a_k}\right)\cdot \vec{v}\,,
\end{equation}
with $\sigma$ being the sigmoid function. This sigmoid function has the effect that the neural network automatically can single out one $x_k$. 

We use an Adam optimizer with a learning rate of $10^{-2}$ if not otherwise stated. We train our network for 50.000 steps, where one step corresponds to one run with a batch size of 2000. We alternate the training between our $L$ and $M$ networks.

\subsection{Harmonic Oscillator}
We start with the example of the harmonic oscillator. The Hamiltonian is $H=\frac{1}{2}p^2 + \frac{\omega^2}{2}q^2$. The equation of motion we want to learn are 

\begin{equation} \begin{split}
	\dot{q} = p\,,\qquad \qquad \dot{p}	= -\omega^2 q\,,
 \end{split} \end{equation}
where we use $\omega =2$. In general, one can also build networks to find the Lax pair for arbitrary $\omega$, but this leads more cumbersome networks, equations and sampling processes.

We sample $10^5$ data points with $p,q \approx \mathcal{N}\left(0,2\right)$ and compute the time derivatives using the equations of motion.

As our neural network ansatz for $L$, $M$ we use 
\begin{equation}
\begin{split} 
	&L_{ij}\left(p,q\right)= a_{ij} + b_{ij} q +c_{ij} p\,,\\
	&M_{ij}\left(p,q\right)= d_{ij} + e_{ij} q + f_{ij} p\,,
\end{split}
\end{equation}
so only polynomials up to first order and $i,j=\lbrace 1, 2\rbrace$. As a general rule, we look at the powers within the equations of motion and use the powers as guidance for an ansatz. Additionally, we restrict the sum over the absolute values of $b_{ij}$ and $c_{ij}$ (to avoid mode collapse for any of them):
\begin{equation} \begin{split}
	\sum_{i,j=1}^2\left|b_{ij}\right|\geq \frac{1}{2}\qquad \text{and}\qquad\sum_{i,j=1}^2\left|c_{ij}\right|\geq \frac{1}{2}\,.
 \end{split} \end{equation}
An example for the Lax pair we find with this ansatz are
\begin{equation} \begin{split}
	L=\left(\begin{matrix}0.437 ~q & - 0.073~ p\\- 0.666~ p & - 0.437 ~q\end{matrix}\right)\,, \qquad
	M=\left(\begin{matrix}0.001 & 0.329\\-3.043 & -0.001\end{matrix}\right)\,,
 \end{split} \end{equation}
where we did not allow linear combinations of the differential equations. This corresponds to a solution of the second type described in~\eqref{eq:HarmonicStandardSolution}. A quick check shows that both sides of the Lax pair condition (\ref{eq:LaxPairCondition}) match accurately
\begin{equation} \begin{split}
	\frac{dL}{dt}=\left(\begin{matrix}0.437~ \dot{q} & - 0.073 ~\dot{p}\\- 0.666~ \dot{p} & - 0.437 ~\dot{q}\end{matrix}\right) =  
	\left(\begin{matrix} 0.441~ p &  0.288~ q\\  2.660~ q &- 0.441~ p\end{matrix}\right) = \left[L,M\right]~.
 \end{split} \end{equation}
We check that $\tr L^2\sim H$:
\begin{equation} \begin{split}
	L^2 = \left(\begin{matrix}0.048618 p^{2} + 0.190969 q^{2} & 0\\0 & 0.048618 p^{2} + 0.190969 q^{2}\end{matrix}\right) \Rightarrow \tr L^2 \approx 0.2~H\,.
 \end{split} \end{equation}
Therefore, we can see that our method is perfectly able to reproduce the Lax pair in the literature.

We train also for the corresponding R-matrices and find as a result
\begin{equation}
	r_{12} = \left(\begin{matrix}
		0&0.92 \\
		-1&0
	\end{matrix}\right)\otimes
	\left(\begin{matrix}
		0&0.92- \frac{0.42}{p} \\
		-1&0
	\end{matrix}\right)
	-\left(\begin{matrix}
		0&0.92- \frac{0.42}{p} \\
		-1&0
	\end{matrix}\right)\otimes
	\left(\begin{matrix}
		0&0.92 \\
		-1&0
	\end{matrix}\right)\,,
\end{equation}
which solves perfectly Equation~\eqref{eq:rmatrixcondition}.

\subsection{Korteweg–de Vries equation}
The second system we consider is the Korteweg–de Vries equation:
\begin{equation} \begin{split}
\label{eq:kdv}
	\dot{\phi} + \phi^{'''} + 6 \phi \phi^{'} = 0\,.
 \end{split} \end{equation}
It describes the behaviour of waves in the shallow water, and is one of the easiest realizations of a classical field theory, and therefore a good starting point for us. A solution of this equations is known to be of the form:
\begin{equation} \begin{split}
	\phi(x,t)= \frac{c}{2} \text{sech}^2\left(\frac{\sqrt{c}}{2}\left(x-c t -a\right)\right)\,,\qquad \text{with } a,c \in \mathbb{R}\,.
 \end{split} \end{equation}
For the sampling process we draw $10^5$ samples from $\phi\,,~\phi^{'}\,,~\phi^{''}\,,~\phi^{'''} \approx \mathcal{N}\left(0,2\right)$ and compute $\dot{\phi}$ using equation \ref{eq:kdv}. 

The ansatz is again determined by the structure of the equation. While we expect the $A_x$ matrix to be linear in $\phi\left(x\right)$, the matrix $A_t$ should be a polynomial up to second order in $\phi\left(x\right)$, $\phi^{'}\left(x\right)$ and $\phi^{''}\left(x\right)$. Due to the fact that we take the derivative $\frac{d}{dx}$ once we do not have to include terms including $\phi^{'''}\left(x\right)$. Using a $2\times2$-matrix as an ansatz, we find:
\begin{equation} \begin{split}
	A_x&=\left(\begin{matrix}1.6 \phi - 0.2 & -0.8\\0.2 & 1.7 \phi + 0.2\end{matrix}\right)\,,\\
	A_t&=\left(\begin{matrix}- 4.9 \phi^{2} - 1.6 \phi^{''} - 0.1 & 0.1 \phi^{2} - 0.3\\0.1 & - 5.0 \phi^{2} - 1.7 \phi^{''} + 0.1\end{matrix}\right)
\,,
 \end{split} \end{equation}
where we rounded entries to $10^{-1}$. If we are interested in an analytical solution one could drop the bias terms, which does not change the solution but makes the solution more familiar.
A quick check shows, that this pair $A_x$, $A_t$ fulfills the KdV-equation:
\begin{equation}
\begin{split} 
	\left(\begin{matrix}1.6 \dot{\phi}  & 0\\
	0 & 1.7 \dot{\phi} \end{matrix}\right)
	-\left(\begin{matrix} - 9.8 \phi \phi^{'} - 1.6 \phi^{'''}  & 0.2 \phi\phi^{'} \\
	0 & - 10.0 \phi\phi^{'} - 1.7 \phi^{'''}\end{matrix}\right) 
	+ \mathcal{O}\left(0.1\right)\approx 0\,.
\end{split}
\end{equation}
but this rather different then in the literature, where the $A_x$-matrix only has off diagonal entries.  Therefore, we started a second run forcing the neural network to find at least one off-diagonal element by demanding that in the $\loss_{\text{MC}}$ only off-diagonal elements go in. This leads to the following results

\begin{equation} \begin{split}
	A_x=&\left(\begin{matrix}- 1.7 \phi & 1.7 \phi + 1.0\\1.7 \phi + 1.0 & - 1.7 \phi\end{matrix}\right)\,, \\
	A_t=&\left(\begin{matrix}5.0 \phi^{2} + 1.7 \phi^{''} & - 5.0 \phi^{2} - 1.7 \phi^{''} - 0.5\\- 5.0 \phi^{2} - 1.7 \phi^{''} - 0.5 & 5.0 \phi^{2} + 1.7 \phi^{''}\end{matrix}\right)
\,,
 \end{split} \end{equation}
which is now much closer to the literature and again provides a pair which fulfills the required conditions:
\begin{equation}
\begin{split} 
	\frac{\partial A_x}{\partial t} -	\frac{\partial A_t}{\partial x} +\left[A_x,A_t\right] = 
	\left(\begin{matrix}- 1.7 \dot{\phi} & 1.7 \dot{\phi} \\1.7 \dot{\phi} & - 1.7 \dot{\phi}\end{matrix}\right)-
	\left(\begin{matrix}10.0 \phi^{'}\phi + 1.7 \phi^{'''} & - 10.0 \phi^{'}\phi - 1.7 \phi^{'''} \\- 10.0\phi^{'}\phi - 1.7 \phi^{'''}  & 10.0 \phi^{'}\phi + 1.7 \phi^{'''}\end{matrix}\right) + \mathcal{O}\left(0.1\right)\approx 0.
\end{split}
\end{equation} Additionally this shows the flexibility of this method which can be adapted to single out solutions which are close to a particular desired form. Further small generalizations of our network such as demanding equality between different prefactors we leave for the future.

\subsection{Heisenberg-magnet}
Let us start with a much more sophisticated problem: the Heisenberg model for a ferromagnet with an SO(3) symmetry group. The Hamiltonian of the problem reads:
\begin{equation} \begin{split}
	H=\frac{1}{2}\int\,dx ~\vec{S}^2\left(x\right),
 \end{split} \end{equation}
where $\vec{S} \in S^2$ and the components satisfy as an additional constraint
\begin{equation} \begin{split}
	\lbrace S_a\left(x\right),S_b\left(y\right)\rbrace = \epsilon_{abc}S_c\left(x\right)\delta\left(x-y\right)\,.
 \end{split} \end{equation}
The equations of motion read
\begin{equation} \begin{split}
	\dot{\vec{S}} = \lbrace H, \vec{S}\rbrace = - \vec{S}(x)\times \vec{S}{}^{''}(x)\,.
 \end{split} \end{equation}
As one can see now we have the three input variables $S_i$ as well as three derivatives of them as input values (the second derivative is again not irrelevant due to the degree of derivatives in the differential equation). We know that we have an underlying SU(2) structure here which we utilize in our ansatz for the Lax connection. It should be invariant under SU(2) transformation and we use as ansatz for $A_x$:
\begin{equation}
	A_x=a~\vec{\sigma}\vec{S} + B \,,\qquad\text{with}\quad a \in \mathbb{C}\,, B \in \mathbb{C}^{(2\times 2)}\,.
\end{equation}
For $A_t$, we use a polynomial ansatz:
\begin{equation}
	A_t = C ~S_i+D ~S_i^{'}+E~ S_i~S_j+F~ S_i~S_j^{'}+G~ S_i^{'}~S_j^{'}\,,\qquad\text{with}\quad C\,,~D\,,~E\,,~F\,,~G~ \in \mathbb{C}^{(2\times 2)}\,.
\end{equation}
Using this ansatz (something which looks reasonable) we find the following formulae for the Lax connection $A_x$ and $A_t$:
\begin{equation} \begin{split}
	A_x=& - ~\text{i}~ \vec{\sigma}\vec{S} +0.3 \left(\begin{array}{cc}
	1 &0\\
	0& 1	
	\end{array}\right),\\\
	A_t=&\left(\begin{array}{cc}
	2 ~\text{i}~ S_z &2 ~\text{i}~ S_x +2 S_y\\
	2 ~\text{i}~ S_x-2 S_y& - ~\text{i}~ S_z
	\end{array}\right)\\
	&+\left(\begin{array}{cc}
	 ~\text{i}~ S^{'}_yS^{}_x- ~\text{i}~ S^{'}_xS^{}_y &- S^{'}_zS_x + S^{'}_xS_z +~\text{i}~ (S^{'}_zS_y-S^{'}_y S_x)\\
	+S^{'}_zS_x - S^{'}_xS_z+~\text{i}~ (S^{'}_zS_y-S^{'}_y S_x)& -~\text{i}~ S^{'}_yS^{}_x+~\text{i}~S^{'}_xS^{}_y
	\end{array}\right)\\
	=& 2~\text{i}~ \vec{\sigma}\vec{S}+~\text{i}~\epsilon_{ijk}\sigma_i S_j S^{'}_k~,
 \end{split} \end{equation}
where we rounded at the least shown digit.
This is perfectly in line with the known literature. Here, we can see the power and the limitations of the technique: It is able to find the right terms for the matrix $A_t$, but we also have to use our physical intuition to restrict $A_x$ in our ansatz.

\subsection{Non-linear sigma models}
Another important class of integrable field theory models is given by non-linear sigma models in two spacetime dimensions which we want to briefly discuss now. We are interested in $O(N)$ linear sigma-models where the fields are living on $S^{N-1}.$ The Lagrangian for these models can be written as
\begin{equation}
 \mathcal{L} = -\text{Tr}~ \left(J_\mu J^\mu\right),\qquad J_\mu=(\partial_\mu g)g^{-1}~,\qquad \mu=0,1~.
\end{equation}
This system obeys the following equations:
\begin{equation}
 \begin{split}
  \partial_\mu J^\mu=~&0~,\\
  \partial_\mu J_\nu-\partial_\nu J_\mu-[J_\mu,J_\nu]=~&0~.\label{eq:principle}
 \end{split}
\end{equation}

We now study two special cases $N=3$ and $N=4$. The former is related to the Sine-Gordon model which can be seen by appropriately re-writing the equations of motion (cf.~Chapter~6.3 of~\cite{doi:10.1142/4678} and~\cite{Pohlmeyer:1975nb}); whereas the latter is related to the principal chiral model.

\subsubsection*{Sine-Gordon equation}
The associated Sine-Gordon equation is given by
\begin{equation} \begin{split}
	\frac{\partial^2}{\partial x\partial t} \phi - \sin\left(\phi\right) = 0\,.
 \end{split} \end{equation}
We are interested in finding a Lax pair associated to this equation of motion. A known solution for the Sine-Gordon equation is:
\begin{equation} \begin{split}
	 \phi\left(x,t\right)= 4~\arctan\left(\exp\left(\gamma \left(x+t-v~(x-t)\right)+\delta\right)\right)\,,\qquad\text{with\,} \gamma^2 =\frac{1}{1-v^2}\,,
 \end{split} \end{equation}
where we sampled $v$ from a uniform distribution ${\cal U}(-0.9,0.9)$ and $\delta$ from the normal distribution $\mathcal{N}\left(-1,1\right)$.
When looking at the differential equation it is rather obvious that our formula might depend on sine or cosine, while $A_x$ only depends on the first spatial derivative of $\phi$
\begin{equation} \begin{split}
	A_{x,ij}&= a_{ij} \phi^{'} + b_{ij}\,,\\
	A_{t,ij}&= c_{ij}\phi +d_{ij} \frac{\partial \phi}{\partial x} +e_{ij}\cos \phi+f_{ij}\sin \phi\,.
 \end{split} \end{equation}
Using this ansatz, again using complex coefficients, we find
\begin{equation} \begin{split}
	A_x=&\left(\begin{matrix}0.5 \phi^{'} - 0.7 - 0.6 ~\text{i}~ & 0.9 \phi^{'} + 0.2 ~\text{i}~ \phi^{'}\\- 0.5 \phi^{'} + 0.1 ~\text{i}~ \phi^{'} + 0.8 + 0.4 ~\text{i}~& - 0.5 \phi^{'} + 0.7 + 0.6 ~\text{i}~\end{matrix}\right)\\
	A_t=&\left(\begin{matrix}- 0.2 \cos \phi + 0.2 ~\text{i}~ \cos \phi + 0.2 \sin \phi - 0.2 ~\text{i}~ \sin \phi & 0.5 \sin \phi - 0.2 ~\text{i}~ \sin \phi\\0.2 \cos \phi - 0.2 ~\text{i}~ \cos \phi & 0.2 \cos \phi - 0.2 ~\text{i}~ \cos \phi - 0.2 \sin \phi + 0.2 ~\text{i}~ \sin \phi\end{matrix}\right)\,.
 \end{split} \end{equation}
We find that the condition for the Lax connection is satisfied
\begin{equation} \begin{split}
\frac{dA_x}{dt}&=\left(\begin{matrix}0.5 \frac{d^2\phi}{dxdt}  & 0.9 \frac{d^2\phi}{dxdt} + 0.2 ~\text{i}~ \frac{d^2\phi}{dxdt}\\- 0.5 \frac{d^2\phi}{dxdt} + 0.1 ~\text{i}~ \frac{d^2\phi}{dxdt} & - 0.5 \frac{d^2\phi}{dxdt} ~\text{i}~\end{matrix}\right)=\\
&=\left(\begin{matrix} 0.5 \sin \phi &0.9 \sin \phi + 0.2 ~\text{i}~ \sin \phi \\
	-0.5 \sin \phi + 0.1 ~\text{i}~ \sin \phi  & - 0.5 \sin \phi \end{matrix}\right)=\frac{dA_t}{dx}+\left[A_x,A_t\right]\,.
 \end{split} \end{equation}
As we can see, the equation is only fulfilled iff the Sine-Gordon equation is fulfilled as well.

\subsubsection*{Principal chiral model}

For the principal chiral model we used again the ansatz with a proportionality to the Pauli matrices (i.e.~re-writing $J=\vec{\sigma}\vec{J}$):
\begin{equation} \begin{split}
	A_x=a~\vec{\sigma}\vec{J}_x+b~\vec{\sigma}\vec{J}_t\,,\qquad\text{with}\quad a,b \in \mathbb{C}\,,
 \end{split} \end{equation}
whereas the matrix $A_t$ is unconstrained. We sample all variables using $\mathcal{N}(0,2)$ for the sampling and the equations of motion (cf.~Equation~\eqref{eq:principle}) for $10^6$ samples. The matrices fulfill the Lax-constraints:
\begin{equation} \begin{split}
	A_x=&\left(0.295+~\text{i}~ 0.205\right)\vec{\sigma}\vec{J}_x+\left(-0.512+~\text{i}~ 0.165\right)\vec{\sigma}\vec{J}_t\,, \\
	A_t=&\left(0.515-~\text{i}~ 0.162\right)\vec{\sigma}\vec{J}_x+\left(-0.283-~\text{i}~0.193\right)\vec{\sigma}\vec{J}_t\,,
 \end{split} \end{equation}
where the matrices have perfectly the shape of the Pauli-matrices. A quick check shows the consistency of our solution with the equations of motion:
\begin{equation} \begin{split}
&	\dot{A}_x - {A}_t^{'}+\left[A_x,A_t \right]= \\
&\left(0.295+~\text{i}~ 0.205\right)\vec{\sigma}\dot{\vec{J}}_x+\left(-0.512+~\text{i}~ 0.165\right)\vec{\sigma}\dot{\vec{J}}_t
    + \left(0.515-~\text{i}~ 0.162\right)\vec{\sigma}\vec{J}_x{}^{'}-\left(0.283+~\text{i}~0.193\right)\vec{\sigma}\vec{J}_t{}^{'}\\
    &-\left[\left(0.295+~\text{i}~ 0.205\right)~\left(0.283+~\text{i}~0.193\right)+\left(0.515-~\text{i}~ 0.162\right)~\left(-0.512+~\text{i}~ 0.165\right)\right]~\text{i}~2~\epsilon^{abc}J_t^bJ_x^c\\
    \approx &\left(-0.512+~\text{i}~ 0.165\right)\vec{\sigma}\left( \dot{\vec{J}}_t-\vec{J}_x{}^{'}\right)\\&\qquad+
    		\left(0.283+~\text{i}~0.193\right)\vec{\sigma}\left(\dot{\vec{J}}_x-\vec{J}_t{}^{'}\right)-2~\left(0.283 +~\text{i}~ 0.193\right)~\epsilon^{abc}\sigma^a J_t^bJ_x^c~,
 \end{split} \end{equation}
where we used the standard commutation relations $[\sigma^a,\sigma^b]=\epsilon^{abc}\sigma^c.$ The Lax connection also can be matched with the solution presented in~\cite{Torrielli:2016ufi}.
\section{Integrable vs Non-integrable systems}
\label{sec:interableornot}
In this section we focus on perturbations within an integrable theory. We use the harmonic oscillator and the Heisenberg models as the unperturbed system respectively and introduce one integrable and one non-integrable perturbation controlled by the parameter $\epsilon$. When increasing $\epsilon$, we check whether the neural network is able to compensate the effect of the perturbation within the Lax pair.
\begin{figure}
\centering
        \includegraphics[scale=.8]{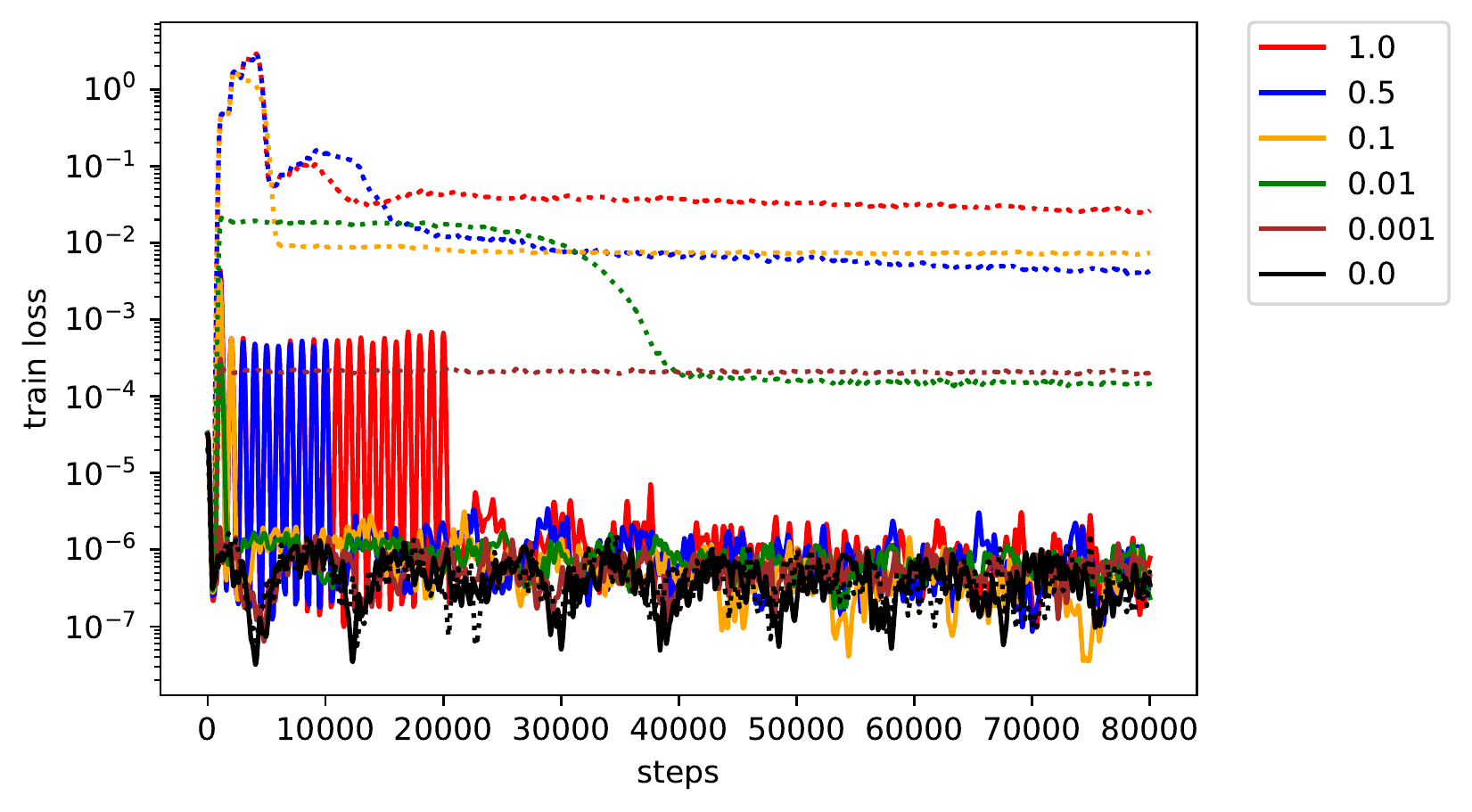}
	\includegraphics[scale=.8]{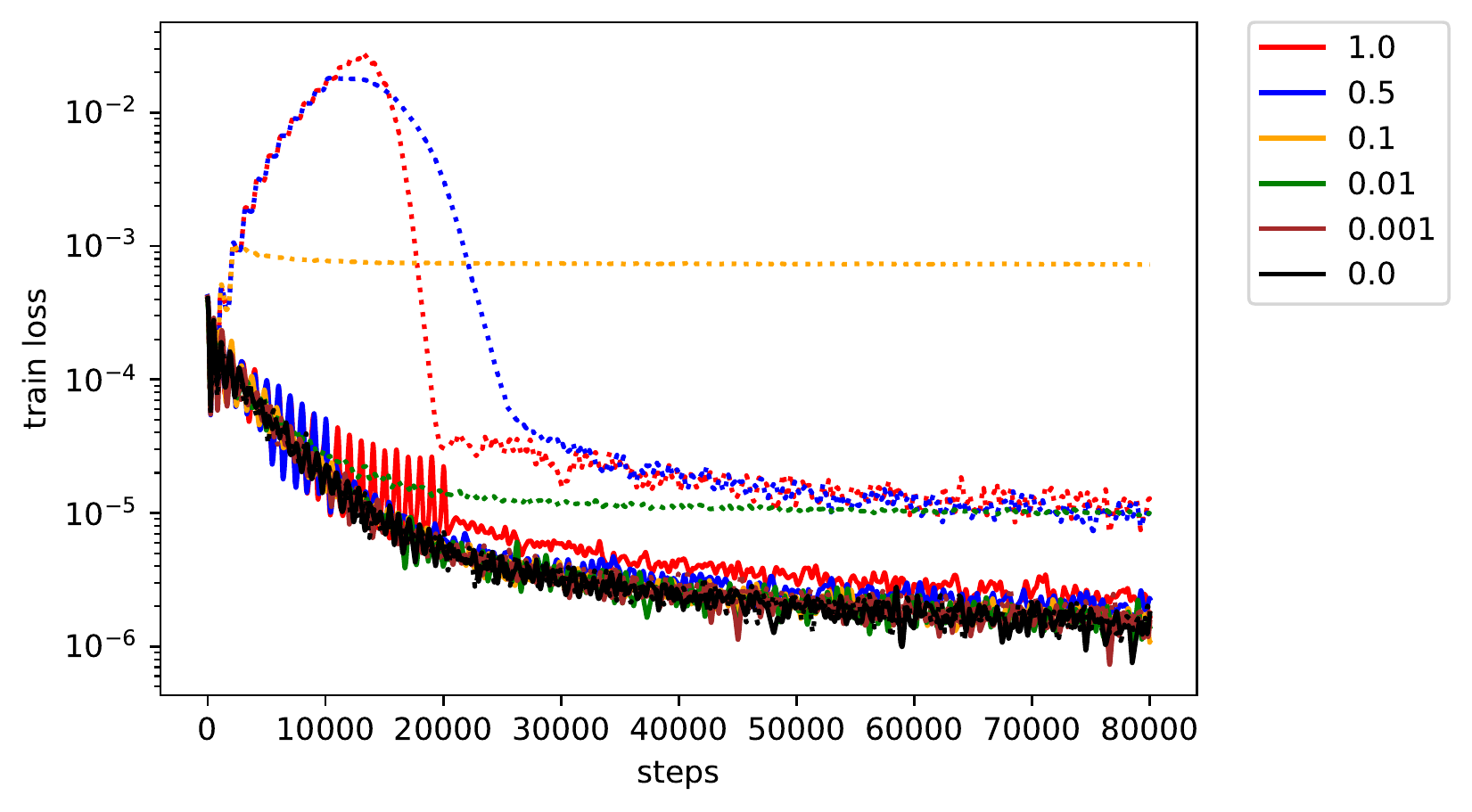}

	\caption{We show the comparison between an integrable (solid lines) and a non-integrable (dotted lines) perturbation for the two-dimensional harmonic oscillator and the Heisenberg model. The upper plot shows the two-dimensional harmonic oscillator and the lower one the Heisenberg model. In both cases, the perturbation grows linearly over time with $\epsilon_{\rm{step}}=0.05$. In the case of the integrable perturbation, the neural network is able to compensate this effect and adapt the Lax pair. Therefore, after some time, the training error decreases. For the non-integrable perturbation this is not the case. Here, the training error grows linearly in time, and therefore, proportional to the perturbation (note the log scale of the y-axis). The steps in the training loss at the beginning for large values of $\epsilon$ correspond to the times when the perturbation is increased.}	\label{fig:comparison}
\end{figure}

\subsection{Two-dimensional Harmonic Oscillator}

We start with the harmonic oscillator in two dimension with a quadratic perturbation and a fourth order perturbation:
\begin{equation}
	\begin{split} 
		H_1 =& \frac{1}{2}\left(p_x^2+p_y^2+x^2+y^2+\epsilon x y\right)\,,\\
		H_2 =& \frac{1}{2}\left(p_x^2+p_y^2+x^2+y^2+\epsilon x^2 y^2\right)\,.
	\end{split}
\end{equation}  

$\epsilon$ is the parameter which scales the perturbation. $H_1$ describes a coupled harmonic oscillator and is therefore integrable, whereas the perturbation in $H_2$ is not integrable.\\
The neural network for $L$ is described in Section~\ref{sec:experiments}, and for $M$ we use a neural network with 2 hidden layers with 200 neurons each and tanh activation on the hidden layers. For each $\epsilon$ we are sampling a new data set with 100000 points. We draw $x$, $y$, $p_x$ and $p_y$ from $\mathcal{N}\left(0,1\right)$, and compute $\dot{x}$, $\dot{y}$, $\dot{p}_x$ and $\dot{p}_y$ using:
\begin{equation}
\begin{split} 
	&\dot{x}= p_x\,,\qquad \dot{y}= p_y\,,\qquad   \dot{p}_x= -x-\epsilon y\,,\qquad  \dot{p}_y= -y-\epsilon x~,\\
&\dot{x}= p_x\,,\qquad \dot{y}= p_y\,,\qquad   \dot{p}_x= -x-\epsilon ~x~y^2\,,\qquad  \dot{p}_y= -y-\epsilon~ y~x^2	\,.
\end{split}
\end{equation}
We trained the neural network for 80000 steps, while we  increased $\epsilon$ 
every 1000 steps for $0.05$ till it reaches $\epsilon_{\rm{final}}$ using the Adam optimizer with learning rate $10^{-3}.$ We trained 
for $\epsilon_{\rm final}=\left(0,0.001,0.01,0.1,0.5,1 \right)$. The evolution 
of the loss can be seen in Figure \ref{fig:comparison} (top) which clearly shows 
that the system can adapt to the integrable perturbation. However, for the 
non-integrable perturbation the network does no longer converge, 
i.e.~hierarchically larger loss values are encountered. Note that such an 
increased loss value is only a hint for a non-integrable perturbation but this 
is not a strict proof as this can be due to a too simple ansatz for the Lax pair 
network. We show the median of 10 runs for each $\epsilon$ and perturbation type 
and the curves are smoothed with a Gaussian filter with $\sigma=10$.

\subsection{Heisenberg-Model}
Finally, we conclude with the Heisenberg model and compare again two perturbations: The integrable perturbation is known from the Landau-Lifschitz equation ($S\wedge JS$):
\begin{equation}
\label{Heisenberg_int}
	\begin{split} 
	\dot{\vec{S}} =- \vec{S}(x)\times \vec{S}{}^{''}(x) - \vec{S} \times J \vec{S}\qquad \text{with } J = \begin{pmatrix}
	-\epsilon & 0 & 0\\
		0 & \epsilon & 0\\
			0 & 0 & 0
	\end{pmatrix}
	\,.
	\end{split}
\end{equation} 
The second example contains as well a perturbation in quadratic order, but now with the structure $ S J S$:
\begin{equation}
\label{Heisenberg_non_int}
	\begin{split} 
	\dot{\vec{S}} =- \vec{S}(x)\times \vec{S}{}^{''}(x)+ \begin{pmatrix}
	\epsilon\, S_x^2 \\
		  -\epsilon\, S_y^2 \\
			0 & 
	\end{pmatrix}
	\,.
	\end{split}
\end{equation}

We start with a correctly initialized neural network for the unperturbed Heisenberg model, and then turn on the perturbation with small steps of $0.05$ to enable the neural network to adapt to the perturbation until we reach the target value for the perturbation. Note, that we used a neural network with output dimensions $4\times 4$ instead of $2\times 2$ to find a suitable Lax-pair.

The neural network for $L$ is described in Section~\ref{sec:experiments}, and 
for $M$ we use a neural network with 2 hidden layers with 200 neurons each and 
tanh activation on the hidden layers. For each $\epsilon$ we are sampling a new 
data set with 100000 points. We draw $S$ and $S^{''}$ from 
$\mathcal{N}\left(0,1\right)$, and compute $\dot{S}$ 
using~\eqref{Heisenberg_int} and~\eqref{Heisenberg_non_int} respectively.

We trained the neural network for 80000 steps, while we  increased $\epsilon$ every 1000 steps until it reaches $\epsilon_{\rm{final}}$ using Adam with learning rate $10^{-3}.$. We trained for $\epsilon_{\rm final}=\left(0,0.001,0.01,0.1,0.5,1 \right)$. The evolution of the loss can be seen in Figure \ref{fig:comparison} which clearly shows that the system can adapt to the integrable perturbation. The neural network for the non-integrable perturbation converge to a solution with larger loss compared to the integrable perturbation and compared to the unperturbed system. Hence we are able to single out the integrable perturbation of the Heisenberg model~\eqref{Heisenberg_int} (see for example the perturbation with $\epsilon = 0.1$). We plot the median of 10 runs for each $\epsilon$ and perturbation type. The curve is smoothed with a Gaussian filter with $\sigma=100$.

\section{Conclusions}
\label{sec:conclusions}
A natural question for every dynamical system is whether it is integrable or not and which symmetries are conserved. Our method provides the basis to search for these structures given the equations of motions of a system {\it ex machina}.

This method enables us now to understand systematically without relying on human intuition to determine whether a system is integrable. In addition, the Lax pair and connection point us at the conserved symmetries of a system.

We have tested our method on known examples in the literature and were able to find associated Lax pairs and connections. Due to the ambiguities in the functional form of Lax pairs, our algorithm identified Lax pairs which we previously had not encountered in the literature but could retrospectively verify. 

In addition, as shown in the simple case of a (perturbed) harmonic oscillator and the Heisenberg model, by writing down (automatically) perturbations to a known integrable model and then by appling our method to this perturbed system, one has a handle on determining whether the system remains integrable. It will be very interesting to apply these methods in the context of perturbations to ${\cal N}=4$ super Yang-Mills theory which  is known to be integrable (in the planar limit) as pioneered in~\cite{Beisert:2003yb} (cf.~\cite{Beisert:2010jr} for an extensive review). By applying our method to perturbations of the ${\cal N}=4$ action to determine under which ones the system remains integrable.

Another interesting future application is to explore the potential connection between integrability and infinite symmetries of CFTs on the celestial sphere such as BMS symmetries (see for instance~\cite{Guevara:2021abz} for recent work on identifying these symmetries). In this context, it seems relevant to study the integrability in sub-sectors of a quantum field theory which is not integrable in general. With our method we can search for Lax pairs associated to particular sub-sectors of the general solution by using only such samples for training (e.g.~we have obtained the Lax pair of the harmonic oscillator from data associated to a single frequency).

We hope to return to such interesting questions on the integrability of systems in the future.

\section*{Acknowledgments}
We would like to thank David Berman and Saskia Demulder for discussions.
The work of D.L. is supported by the Excellence Cluster `Origins'.

\bibliography{NewBib} 
\bibliographystyle{JHEP}

\end{document}